\documentclass[pra, twocolumn,amsmath, notitlepage,showpacs]{revtex4-1}
\usepackage{bm,bbm,hyperref,graphicx}
\usepackage[utf8]{inputenc} 

\hypersetup{
    colorlinks = true,
    citecolor=blue,
    linkcolor = red,
    anchorcolor = red,
    citecolor = blue,
    filecolor = red,
    pagecolor = red,
    urlcolor = red,
    }
\begin{document}	
	\title{The qubit reveals a qubit-cavity system}
	\author{Sreenath K.~Manikandan}
	\affiliation{School of Physics, IISER TVM, CET Campus, Thiruvananthapuram, Kerala, India 695016}
	\author{Vinayak Jagadish}
	\email[email: ]{vinayak@iisertvm.ac.in}
	\affiliation{School of Physics, IISER TVM, CET Campus, Thiruvananthapuram, Kerala, India 695016}
	\author{Anil Shaji}
	\affiliation{School of Physics, IISER TVM, CET Campus, Thiruvananthapuram, Kerala, India 695016}	
	\begin{abstract}
		We show that the coupling between a qubit and a single mode cavity can be estimated from the process tomography data for the qubit alone. All the parameters of the coupling Hamiltonian between the qubit and the cavity mode can be obtained from observations on the qubit. We also show that the variance matrix and the photon number statistics of the single mode cavity can be reconstructed. Our results provide an alternate means of obtaining the coupling and reconstructing the state of the cavity mode in comparison with the techniques presently used. 
	\end{abstract}
	\maketitle
	\section{Introduction}
	A single qubit interacting with a single quantized radiation mode is the prototypical problem in quantum electrodynamics. The paradigm of cavity quantum electrodynamics (Cavity QED) \cite{haroche_exploring_2006} made this experimentally realizable wherein the interaction between a qubit and an optical or microwave cavity with a single  mode can be studied in a controlled manner. Micro-resonators with high Q-factors~\cite{spillane_ultrahigh-$q$_2005} are widely used to prepare single mode cavities.~\cite{langford_circuit_2013}
In recent years, prescriptions have been provided to characterize the cavity and also to prepare cavities in desired initial states~\cite{franca_santos_conditional_2001,davidovich_mesoscopic_1996}.

A relevant and often unknown parameter in a typical cavity QED experiment is the coupling between the qubit and the cavity. Calculating the coupling from first principles~\cite{kimble_strong_1998} using parameters such as the atomic dipole moment, field amplitude and the volume of the cavity~\cite{hood_real-time_1998} is typically based on several assumptions and idealizations, necessitating a direct estimation of the coupling in a real experiment. This is done~\cite{niemczyk_circuit_2010,badolato_deterministic_2005,brune_quantum_1996,brune_lamb_1994} by observing the Rabi oscillations between the normal modes of the system~\cite{thompson_observation_1992,peter_exciton-photon_2005}. The splitting between the levels is proportional to the coupling at resonance, and it is also related to the excitation quanta of the cavity. The experimentally estimated coupling strength may deviate~\cite{badolato_deterministic_2005} from the number predicted by the cavity geometry~\cite{hood_real-time_1998}.	
	
 In this Paper we investigate a method of estimating the qubit-cavity coupling from the observed dynamics of the qubit alone in the strong coupling regime.  A similar question was previously addressed for the case of two coupled qubits~\cite{jordan07a} as well as for a qubit coupled to an $N$ level system~\cite{jagadish_probing_2014} which showed that the coupling could be obtained from the process tomography data of the qubit alone.  We also obtain the variance matrix~\cite{simon_quantum-noise_1994} of the initial quantum state of the cavity. Our recipe can be extended to obtain the higher order moments of the initial state of the cavity as well,  leading to a complete quantum state tomography of the cavity mode~\cite{lvovsky_continuous-variable_2009,miroshnichenko_quantum_2014,smithey_measurement_1993}. While state tomography of the cavity mode has been done~\cite{tufarelli_oscillator_2011,miroshnichenko_quantum_2014,walser_magnetic_1996}, we provide an alternative approach for doing the same. Compared to the available procedures for estimating the coupling~\cite{niemczyk_circuit_2010,badolato_deterministic_2005,brune_quantum_1996,brune_lamb_1994}, our prescription does not demand any prior knowledge on the excitation quanta of the cavity.

\section{A qubit coupled to single cavity mode}

	The state of the qubit is described by the Pauli matrices, ${\:\sigma_{1},\: \sigma_{2},\: \sigma_{3}\:}$ which satisfy the SU(2) Lie algebra \[[\sigma_{i},\sigma_{j}] = 2i\epsilon_{ijk}\sigma_{k}\]
	The single mode cavity has a characteristic frequency $\omega$ and it is described by the bosonic operators $b$ and $b^{\dagger}$ satisfying the usual commutation relation,
	\[[b,b^{\dagger}] = 1\]
	First, we consider the strongly coupled case where the atom and the cavity mode are near resonance. In this limit, the strong interaction between the two is governed by the Jaynes-Cummings~\cite{jaynes_comparison_1963} Hamiltonian. In Section.~\ref{dispersive}, we also consider the case when the qubit and the cavity are weakly interacting~\cite{blais_cavity_2004,schuster_resolving_2007}, which is the off-resonant or dispersive limit of cavity QED. The Jaynes-Cummings Hamiltonian is given by
	\begin{equation} 
		\label{eq:JChamiltonian}
		 H = \frac{a}{2}\sigma_{3} + \omega \bigg( b^{\dagger}b + \frac{1}{2} \bigg) + g\big( b\sigma_{+} + b^{\dagger}\sigma_{-} \big), 
	\end{equation}
where $\sigma_{\pm} = (\sigma_{1} \pm i \sigma_{2})/2$.	Our aim is to estimate the parameters $a$, $\omega$ and $g$ appearing in $H$. We have used atomic units ($\hbar = 1$) in the above expression and the rest of the Paper.
	
	To obtain the  parameters, the qubit is initialized in the following three states,
	\[\rho^{(k)}_{0} = \frac{1}{2}(\openone + \sigma_{k}) \qquad k = 1,2,3\]
	At the moment, we do not worry about the specific initial state for the cavity, even though prescriptions are available to initialize the cavity to a handful of possible states~\cite{franca_santos_conditional_2001,davidovich_mesoscopic_1996}.
	
	The state of the qubit, in the Schr\"{o}dinger picture, at any later time $t$ can be written as
	\[\rho^{(k)}_{t} = \frac{1}{2} \big[ \openone + c^{(k)}_{1}(t)\sigma_{1} + c^{(k)}_{2}(t)\sigma_{2} + c^{(k)}_{3}(t)\sigma_{3} \big] \]
	Here, $\rho^{(k)}_{t}$ is the density operator of the qubit obtained by tracing out the cavity,
	for different initial states of the qubit $\rho^{(k)}_{0}$. 
	
	The dynamics of the qubit is specified by the nine quantities $c^{(k)}_{i}(t)$ which we assume are obtained from a process tomography experiment corresponding to a duration $t$ on the qubit. Notice that $c^{(k)}_{i}(t)$ are obtained as the expectation values of Pauli operators with respect to the specific initial state as
	\begin{equation}
		\label{eq:trace1}
		c^{(k)}_{i}(t) = \langle\sigma_{i}\rangle^{(k)}_{t} = \text{tr}[\rho^{(k)}_{t}\sigma_{i}] = \text{tr}[\eta^{(k)}_{t}\sigma_{i}\otimes \openone_{C}]
	\end{equation}
	Here $\openone_{C}$ is the identity operator acting on the cavity. Since full or partial traces of the identity operator with respect to any state of the cavity is equal to unity, we avoid writing $\openone_{C}$ explicitly in the following discussion. The second trace in Eq.~(\ref{eq:trace1}) is evaluated over the density operator for the combined system $\eta^{(k)}_{t}$. Switching to the Heisenberg picture, where the time dependence is associated with the operators, we notice that
	\[c^{(k)}_{i}(t) = \text{tr}[\eta^{(k)}_{0}\sigma_{i}(t)]\]
	In the above expression, $\eta^{(k)}_{0}$ is the initial state of the qubit and the cavity taken together. Since we have prepared the qubit in a specific initial state, we know that the combined initial state is separable.
\begin{equation}
\label{initalState}
	\eta^{(k)}_{0} = \rho^{(k)}_{0}\otimes\rho^{(c)}_{0}
\end{equation}
Here  $\rho^{(c)}_{0}$ is any arbitrary quantum state of the single mode cavity which can be repeatedly prepared. Later on we show that our treatment can be used to obtain partial information about $\rho^{(c)}_{0}$.

	In the Heisenberg picture, the $n^{\rm th}$ time derivative of $c^{(k)}_{i}(t)$ is,
	\begin{equation}
	\label{deriv}
	\frac{d^{n}}{dt^{n}}c^{(k)}_{i}(t) = \bigg\langle \frac{d^{n}}{dt^{n}}\sigma_{i}(t)\bigg\rangle_{k} = \text{tr}\bigg[\eta^{(k)}_{0}\frac{d^{n}}{dt^{n}}\sigma_{i}(t)\bigg]
	\end{equation}
The derivatives can be computed from the time series data available from process tomography experiments  on the qubit which makes the scheme experimentally realizable. In what follows, we will show that the time derivatives of  $c^{(k)}_{i}(t)$ contain extractable information regarding the Hamiltonian parameters and the initial state of the cavity.

We start with the first derivative of $\sigma_{1}(t)$, 
\begin{eqnarray*}
	\dot{\sigma_{1}}(t) &=&\frac{d}{dt}\sigma_{1}(t) = i[H,e^{iHt}\sigma_{1}(0)e^{-iHt}]\nonumber\\ &=& i e^{iHt}[H,\sigma_{1}(0)]e^{-iHt}\nonumber \\
	& = & -a\sigma_{2}(t)+ig(b-b^{\dagger})\sigma_{3}(t). 
\end{eqnarray*}
Similarly the other two first derivatives can be evaluated as
	\begin{eqnarray*}
		\dot{\sigma_{2}}(t) &=& a\sigma_{1}(t)-g(b+b^{\dagger})\sigma_{3}(t)\\
		\dot{\sigma_{3}}(t) &=& g(b+b^{\dagger})\sigma_{2}(t)-ig(b-b^{\dagger})\sigma_{1}(t)
	\end{eqnarray*}
	
	Using these relations we calculate the first derivative of $c^{(k)}_{i}(t)$ defined in Eqn.~(\ref{deriv}) at $t$ = 0. The initial state of the qubit-cavity system is separable as in Eqn.~(\ref{initalState}) and the trace appearing in Eqn.~(\ref{deriv})  can be decoupled and computed as product of traces over the qubit and the cavity mode separately. We obtain the following relations,
	\begin{eqnarray}
		\label{eq:FD}
		\dot{c}^{(2)}_{1}(0) &=& -a \nonumber \\
		\dot{c}^{(3)}_{1}(0) &=& ig\langle b-b^{\dagger}\rangle \nonumber \\
		\dot{c}^{(3)}_{2}(0) &=& - g\langle b+b^{\dagger}\rangle
	\end{eqnarray}
We notice that the tensor $\dot{c}^{(k)}_{i}(0)$ is anti-symmetric in $i$ and $k$. From Eq.~(\ref{eq:FD}) we obtain, 
	\begin{eqnarray}
		\label{eq:FDinv}
		a & = & - \dot{c}^{(2)}_{1}(0), \nonumber \\
		\langle b-b^{\dagger}\rangle & = & \frac{\dot{c}^{(3)}_{1}(0)}{ig} , \nonumber \\
		\langle b+b^{\dagger}\rangle & = & -\frac{\dot{c}^{(3)}_{2}(0)}{g}.
	\end{eqnarray}
The first derivatives have fixed one Hamiltonian parameter, namely the free evolution parameter, $a$, of the qubit. This is an expected result since we are looking at the dynamics of the qubit system alone. 
	
	In the same spirit, we also ask for the second derivatives of $c^{(k)}_{i}(t)$. The second derivative of $\sigma_{1}(t)$ is
	\begin{eqnarray*}
	\ddot{\sigma_{1}}(t) &=& i[H,\dot{\sigma_{1}}(t)] \\
	 &=& e^{iHt}i[H, i[H,\sigma_{1}(0)]\:] e^{-iHt}, \\
	 & = & -a^{2}\sigma_{1}(t)+ag(b+b^{\dagger})\sigma_{3}(t)-  g^{2}(i(b-b^{\dagger}))^{2}\sigma_{1}(t)  \\
	 && \quad +g\omega(b+b^{\dagger})\sigma_{3}(t)  + ig^{2}(b^{^{2}}-b^{\dagger^{2}})\sigma_{2}(t)
	\end{eqnarray*}
The second derivatives of other Pauli matrices are also computed in a straight forward manner.
	\begin{equation}
	\begin{split}
	\ddot{\sigma_{2}}(t) = -a^{2}\sigma_{2}(t)+iag(b-b^{\dagger})\sigma_{3}(t)-g^{2}(b+b^{\dagger})^{2}\sigma_{2}(t)\\+ig\omega(b-b^{\dagger})\sigma_{3}(t)+ig^{2}(b^{^{2}}-b^{\dagger^{2}})\sigma_{1}(t),\nonumber
	\end{split}
	\end{equation}		
	and
	\begin{eqnarray}
	\ddot{\sigma_{3}}(t) &=& ga(b+b^{\dagger})\sigma_{1}(t)-g^{2}(b+b^{\dagger})^{2}\sigma_{3}(t)\nonumber \\&-&  ig\omega(b-b^{\dagger})\sigma_{2}(t)+iga(b-b^{\dagger})\sigma_{2}(t)\nonumber \\&-& g^{2}(i(b-b^{\dagger}))^{2}\sigma_{3}(t)-g\omega(b+b^{\dagger})\sigma_{1}(t)-2g^{2}\nonumber
	\end{eqnarray}

	We employ Eq.~(\ref{deriv}) again to compute the second derivatives of $c^{(k)}_{i}(t)$. From $\ddot{\sigma_{1}}(t)$ evaluated at $t = 0$,  we obtain the following relations.
	\begin{eqnarray}
	\label{eq:SD1}
		\ddot{c_{1}}^{(1)}(0) &=& -a^{2}-g^{2}\langle (i(b-b^{\dagger}))^{2}\rangle, \nonumber\\
		\ddot{c_{1}}^{(2)}(0) &=& ig^{2}\langle b^{^{2}}-b^{\dagger^{2}}\rangle, \nonumber\\
		\ddot{c_{1}}^{(3)}(0) &=& ag\langle b+b^{\dagger}\rangle + g\omega\langle b+b^{\dagger}\rangle.
	\end{eqnarray}
	Another set of equations are obtained from $\ddot{\sigma_{2}}(t)$ as,
	\begin{eqnarray}
	\label{eq:SD2}
		\ddot{c_{2}}^{(1)}(0) &=& ig^{2}\langle b^{^{2}}-b^{\dagger^{2}}\rangle,\nonumber\\
		\ddot{c_{2}}^{(2)}(0) &=& -a^{2}-g^{2}\langle(b+b^{\dagger})^{2}\rangle, \nonumber\\
		\ddot{c_{2}}^{(3)}(0) &=& iag\langle b-b^{\dagger}\rangle + ig\omega\langle b-b^{\dagger}\rangle.
	\end{eqnarray}
The remaining set of relations are obtained from $\ddot{\sigma_{3}}(t)$: 
	\begin{eqnarray}
		\label{eq:SD3}
		\ddot{c_{3}}^{(1)}(0) &=& ga\langle b+b^{\dagger}\rangle-g\omega\langle b+b^{\dagger}\rangle-2g^{2}, \nonumber\\
		\ddot{c_{3}}^{(2)}(0) &=& -ig\omega\langle b-b^{\dagger}\rangle+iga\langle b-b^{\dagger}\rangle-2g^{2}, \nonumber\\
		\ddot{c_{3}}^{(3)}(0) &=& -g^{2}\langle(b+b^{\dagger})^{2}\rangle-g^{2}\langle(i(b-b^{\dagger}))^{2}\rangle-2g^{2}.  \quad  
	\end{eqnarray}	
Though most of the relations are redundant here, they are useful in checking the consistency of the prescription we provide. Partial information regarding any of the parameters will also help us to decide the resolution of the time series data needed for a demanded accuracy on parameter estimation. 

The second derivatives $\ddot{c_{1}}^{(1)}$, $\ddot{c_{2}}^{(2)}$ and $\ddot{c_{3}}^{(3)}$ evaluated at $t = 0$ can be used to compute the coupling $g$ as
	\begin{equation}
	\label{eq:g}
	g = \sqrt{\dfrac{1}{2} \big[ \ddot{c_{1}}^{(1)}(0)+\ddot{c_{2}}^{(2)}(0)-\ddot{c_{3}}^{(3)}(0) + 2a^{2} \big].}\end{equation}
	We obtain the expectation values $\langle b \pm b^{\dagger} \rangle$ by substituting Eq.~(\ref{eq:g}) into Eq.~(\ref{eq:FDinv}). These expectation values are proportional to the mean values of the $\hat{x}$ and $\hat{p}$ quadratures of the cavity mode with 
	\[\displaystyle\hat{x} = \frac{1}{\sqrt{2}}( b+b^{\dagger}) \quad  {\rm and} \quad \displaystyle \hat{p} = \frac{1}{i\sqrt{2}}( b-b^{\dagger}) . \]    
	If the cavity mode is in a Gaussian state, then the state is fully characterised by the variance matrix defined as 
	\[{\mathcal V}=  \left(
    \begin{array}{cc}
      \langle \hat{X}^2\rangle &\langle\dfrac{1}{2}\lbrace \hat{X},\hat{P}\rbrace\rangle\\
     \langle\dfrac{1}{2}\lbrace \hat{X},\hat{P}\rbrace\rangle &\langle \hat{P}^2\rangle
    \end{array}
    \right), \]
    with  $\hat{X} = \hat{x} -  \langle \hat{x}\rangle$ and  $\hat{P} = \hat{p} -  \langle \hat{p}\rangle$. By inverting a subset of equations from Eqs.~(\ref{eq:SD1}), (\ref{eq:SD2}) and (\ref{eq:SD3}) above and using Eq.~(\ref{eq:FDinv}) we obtain the elements of the variance matrix of the cavity mode as, 
	\begin{eqnarray*}
	{\mathcal V}_{xx} &=&   \langle \hat{X}^2\rangle = \dfrac{1}{2 g^{2}} \big[ -a^{2}-\ddot{c_{2}}^{(2)}(0)-(\dot{c_{2}}^{(3)}(0))^{2} \big]\\
	{\mathcal V}_{pp} &=&  \langle \hat{P}^2\rangle = \dfrac{1}{2g^{2}} \big[ -a^{2}-\ddot{c_{1}}^{(1)}(0)-(\dot{c_{1}}^{(3)}(0))^{2} \big]\\
	{\mathcal V}_{xp} &=& {\mathcal V}_{px} = \dfrac{\langle \lbrace \hat{X},\hat{P}\rbrace \rangle}{2}  =  \dfrac{1}{2 g^{2}} \big[\!\!-\!\ddot c_{1}^{(2)}(0) \! - \!\dot c_{1}^{(3)}(0)\dot c_{2}^{(3)}(0) \big].	\end{eqnarray*}

	\section{Numerical Example}
We have simulated a simple numerical example for the qubit-cavity system in the absence of real experimental data. For this purpose, we assumed the Hamiltonian parameters $a = 1$, $\omega =1$ and $g = 1$. We also assumed that the cavity is initially prepared in a coherent state with $\langle \hat{N}\rangle = 1$. The time series data for different initial preparation of the qubit state $\rho^{(1)}_{0}$, $\rho^{(2)}_{0}$ and $\rho^{(3)}_{0}$ were constructed from the exact Hamiltonian evolution using MATLAB. For numerical evaluations, the number of levels in the quantum oscillator were reduced to four hundred. The data points were taken at discrete points to mimic a real experiment, by evaluating the reduced density matrix for the qubit at those points. An additional noise could also be added to the data points, but no such noise has been added to the time series data we used here.

With the artificial time series data we constructed, we computed the first derivatives $\dot{c_{i}}^{(k)}(0)$ and the second derivatives $\ddot{c_{i}}^{(k)}(0)$ at $t = 0$. The derivatives evaluated at $t = 0$ are only approximates to the true value of the derivatives. They were computed using a finite but small step size $\delta$, which cannot be arbitrary small in a real experiment. Typically, the smallest step size is the time taken by the qubit to pass through the cavity after its initial preparation.

\begin{figure}[!htb]
		\resizebox{8.5 cm}{7 cm}{\includegraphics{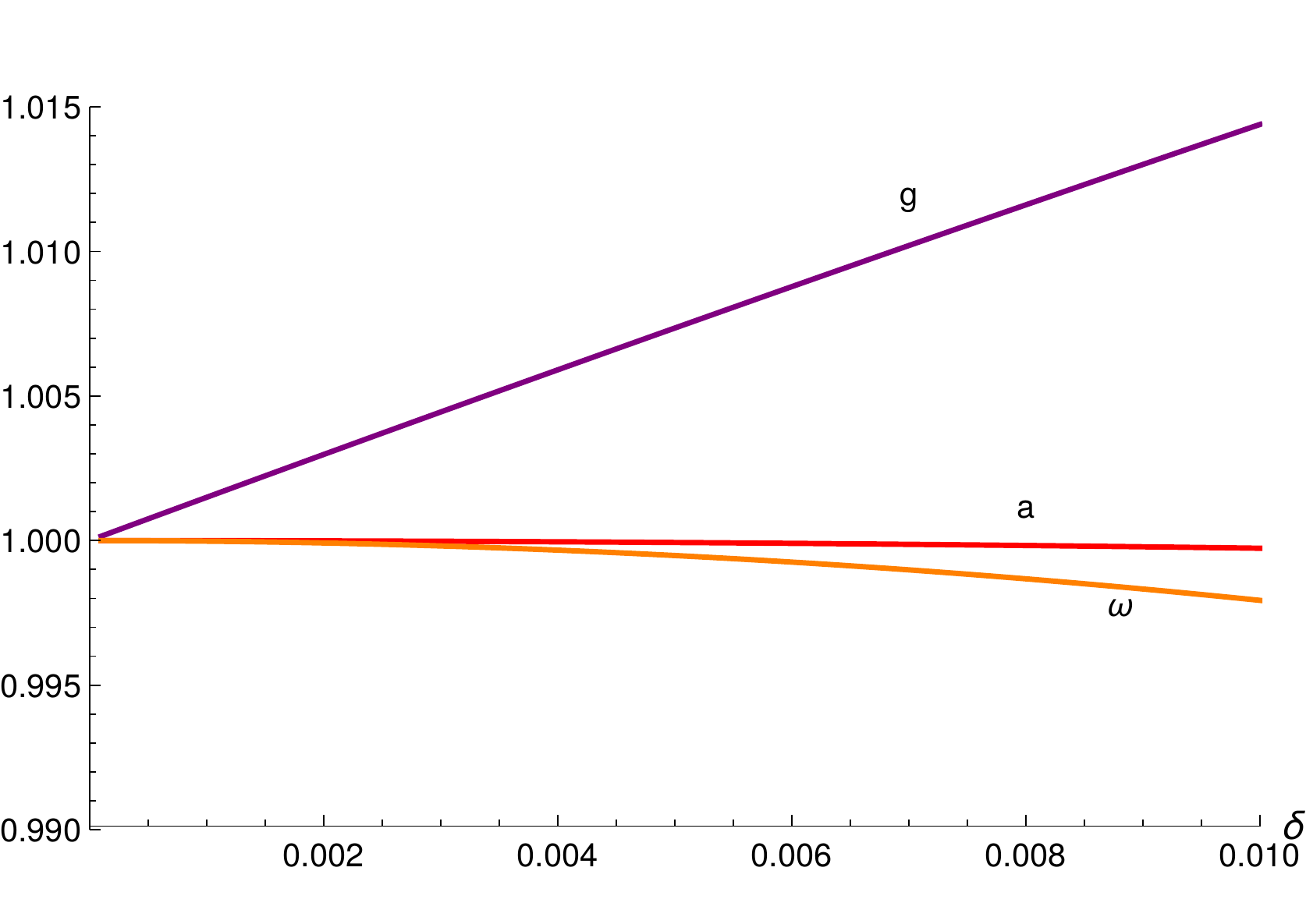}}
		\caption{Reconstructed Hamiltonian parameters as a function of the step size $\delta$ used to compute the derivate. The true value assumed is unity.\label{fig1}}
	\end{figure}

\begin{figure}[!htb]
	\resizebox{8.5cm}{7 cm}{\includegraphics{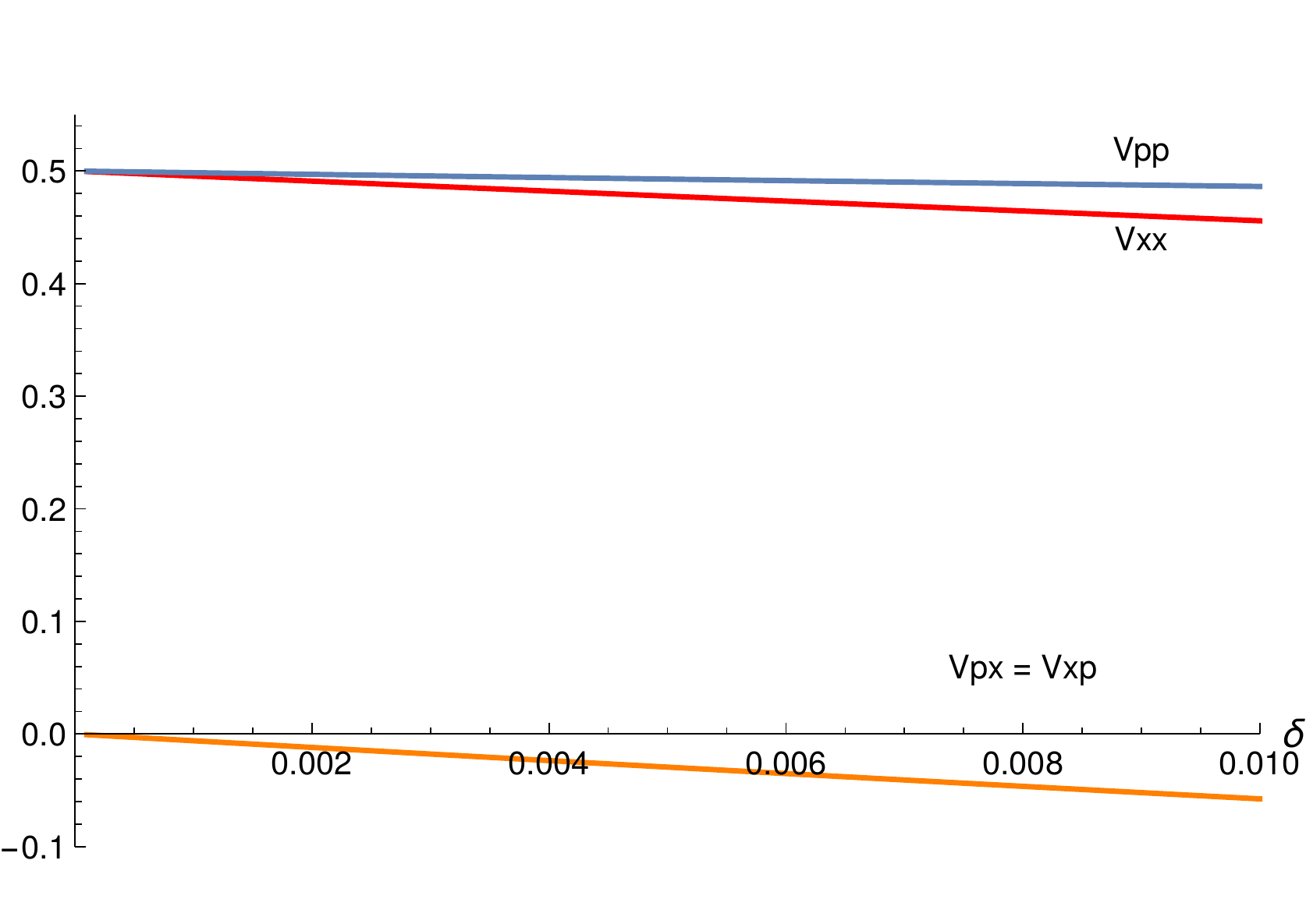}}
	\caption{Reconstructed variance matrix elements for the coherent state with photon number one, for different step sizes $\delta$. \label{fig2}}
\end{figure}

We then used the relations obtained in the previous sections to reconstruct the Hamiltonian parameters as well as the variance matrix of the cavity mode. The reconstructed Hamiltonian parameters are shown in Fig.~\ref{fig1} and the variance matrix elements are shown in Fig.~\ref{fig2}. We observe that the maximum relative error is less than two percentage, for a step size of $0.01$. Notice that the natural timescale in the system, set by the inverse of the Hamiltonian parameters is also equal to one, since we have set all the Hamiltonian parameters equal to one. 

\section{Dispersive Limit \label{dispersive}}

Another regime of interest in Cavity QED, is the dispersive (off-resonant) limit. The system is then described by the following Hamiltonian~\cite{blais_cavity_2004}:
\begin{eqnarray}
H &=& \omega(\hat{N} + 1/2) + \dfrac{a}{2}\sigma_{3}+\dfrac{g^{2}}{\Delta}(\hat{N} + 1/2)\sigma_{3}\nonumber\\ &=& \omega(\hat{N} + 1/2) +\dfrac{\hat{A}}{2}\sigma_{3}
\end{eqnarray}

Here $\Delta$ is the atom cavity detuning given by $\Delta = a-\omega$ and $[H,\hat{A}] = 0$. Note that the $a$ and $g$ appearing in this Hamiltonian is the same as in Eq.(\ref{eq:JChamiltonian}). This weak coupling between the qubit and the cavity has been used to resolve the number states of the cavity~\cite{schuster_resolving_2007}. Here we show that our method can be used to probe the photon number statistics of the cavity by revealing the mean and the variance of this distribution. Here we assume that the parameters $a$ and $g$ have already been estimated using the methods outlined above. 
 
We notice that $\dot{\sigma}_{1}(t) = -\hat{A}\sigma_{2}(t)$ and  $\dot{\sigma}_{2}(t) = \hat{A}\sigma_{1}(t)$. Similarly, the second derivatives are evaluated to obtain  $\ddot{\sigma}_{1}(t) = -\hat{A}^{2}\sigma_{1}(t)$ and $\ddot{\sigma}_{2}(t) = -\hat{A}^{2}\sigma_{2}(t)$. These relations give us the following expressions:
 \begin{eqnarray}
 	\label{eq:weakcoupling1}
 	\dot{c}_{2}^{(1)}(0) &=& \langle \hat{A}\rangle = a+2\dfrac{g^{2}}{\Delta} \bigg( \langle \hat{N} \rangle +\dfrac{1}{2} \bigg) \nonumber \\
 		-\ddot{c}_{1}^{(1)}(0) &=& \langle \hat{A}^{2}\rangle = 4\dfrac{g^{4}}{\Delta^{2}}\langle \hat{N}^{2}\rangle + \bigg( 4\dfrac{g^{4}}{\Delta^{2}}+ 4 a\dfrac{g^{2}}{\Delta} \bigg)\langle \hat{N}\rangle\nonumber\\ &+& a^{2} + \dfrac{g^{4}}{\Delta^{2}}+2a\dfrac{g^{2}}{\Delta}
 \end{eqnarray}
 Eq.~(\ref{eq:weakcoupling1}) can be inverted to obtain 
 \begin{eqnarray*}
 \langle \hat{N}\rangle &=& \dfrac{(\dot{c}_{2}^{1}(0)-a)}{2g^{2}}\Delta - \dfrac{1}{2}\\
 \text{Var}(\hat{N}) &=& \langle \hat{ N}^{2}\rangle - \langle \hat{N}\rangle^{2}\\
&=&  \dfrac{\Delta^{2}}{4g^{4}}\bigg\{ -\ddot{c}_{1}^{(1)}(0) - \bigg[ \bigg( 4\dfrac{g^{4}}{\Delta^{2}} + 4 a\dfrac{g^{2}}{\Delta} \bigg)\langle \hat{N}\rangle \\ 
&& \qquad  + \,  a^{2} + \dfrac{g^{4}}{\Delta^{2}}+2a\dfrac{g^{2}}{\Delta}\bigg]\bigg\} -\langle \hat{N}\rangle^{2}
 \end{eqnarray*}
This procedure of evaluating the higher derivatives of the time series data near $t = 0$ can be extended to obtain the desired higher order moments of the photon number distribution of the initial state of the cavity.
 \section{Conclusion}
In this Paper, we provide a scheme for estimating the Hamiltonian parameters and for obtaining the variance matrix for the quantum state of the cavity by observing the dynamics of a qubit in a cavity-QED system. In addition  to obtaining the variance matrix~\cite{simon_quantum-noise_1994} which completely characterizes Gaussian states of the cavity, our recipe can also be extended to obtain the higher order moments of the initial state of the cavity. The variance matrix of the initial state of the cavity mode is useful for designing quantum information processing protocols using  continuous variable quantum systems. We also studied the dispersive limit,  and we have provided an alternative prescription to resolve the photon number states of the cavity~\cite{schuster_resolving_2007}
by measuring the qubit. The calculations presented here can be repeated for a situation when there are $N$ cavity modes interacting
with a qubit, with characteristic frequencies $\omega_{i}$ and couplings $g_{i}$. Our treatment of evaluating the time
derivatives of the process tomography data up to second order in this case would give us the sum of squares of the couplings $\sum_{i = 1}^{N} |g_{i}|^2$
and no further details about the distribution of the couplings of the $N$ modes to a qubit.
	 \begin{acknowledgments}

S. K. M. acknowledges the support of the Department of Science and Technology, Government of India, through the INSPIRE fellowship scheme (No.~DST/INSPIRE-SHE/IISER-T/2008). A. S. acknowledges the support of the Department of Science and Technology, Government of India, through the Ramanujan Fellowship program (No. SR/S2/RJN- 01/2009).
 \end{acknowledgments}

	\bibliography{paper}
\end{document}